\documentclass[aps,prx,twocolumn,showpacs]{revtex4-1}
\usepackage{mathrsfs}
\usepackage{amsmath}
\usepackage{graphics}
\usepackage{epsfig}
\usepackage{color}
\usepackage{amsfonts,amsmath,mathrsfs}
\usepackage{slashed}
\usepackage{hyperref}
\usepackage{epsfig}
\usepackage{latexsym}
\usepackage{bbm}
\usepackage{amssymb}
\usepackage{graphicx}
\usepackage{dcolumn}
\usepackage{bm}

\begin{document}
\title{Monopoles, confinement and charge localization in the $t$-$J$ model with dilute holes}

\author{Peng Ye$^{1\dag}$ and Qing-Rui Wang$^2$}
\affiliation{$^1$ Perimeter Institute for Theoretical Physics, Waterloo, Ontario,
Canada, N2L 2Y5\\
$^2$ Institute for Advanced Study, Tsinghua University, Beijing, 100084, People's Republic of China}
\begin{abstract}
We present a quantum field theoretic description on the $t$-$J$
model on a square lattice with dilute holes (i.e. near half-filling), based on the compact mutual Chern-Simons gauge theory. We show that, due to the presence of non-perturbative
monopole plasma configuration from the antiferromagnetic background, holons (carrying
electric charge) are linearly confined and strongly localized
even without extrinsic disorder taken into account. Accordingly, the
translation symmetry is spontaneously broken at ground state. Such
an exotic localization is distinct from Anderson localization and
essentially rooted in intrinsic Mott physics of the $t$-$J$ model.
Finally, a finite-temperature phase diagram is proposed. The
metal-insulator transition observed in in-plane resistivity
measurement is identified to a confinement-deconfinement transition
from the perspective of gauge theory. The transition is characterized by the order parameter ``Polyakov-line''.
\end{abstract}

\date{{\small \today}}
\maketitle


\section{Introduction}
A major issue of the single-band $t$-$J$ model on a
square lattice\cite{Anderson87,Lee06}  is how the doped holes (i.e.
Zhang-Rice singlets\cite{ZS} modeling the copper-oxygen
hybridization) interact with spin background eventually producing
plentiful phase diagram of cuprate high-temperature superconductors.
Analytic study on this model is intricate due to the projective
Hilbert space (i.e. two electrons are prohibited to simultaneously
occupy the same site) in which electrons are ``fractionalized'' with
spin-charge separation and emergent gauge degrees of
freedom.\cite{Lee06,Kivelson87,Zou88,Senthil00} These exotic
phenomena open a new window for condensed matter physicists
searching for unconventional quantum states of matter, and also shed
lights on the amazing unified description of fundamental physical laws at different length scales.\cite{Wenbook}

At half-filling (each site is occupied by one electron), the $t$-$J$
model is reduced to quantum Heisenberg model which possesses
antiferromagnetic long range order (AFLRO) and Mott insulating
property.\cite{Chakravarty88,Arovas88PRB,Lee06} At this parent
state, electric charge of electrons is totally frozen out rendering
a pure spin model. Near half-filling where hole concentration is
extremely low, charge localization and insulating ground state have
been read out either directly or indirectly from various types of
experimental justification.\cite{Kastner98,A1,A2,A3}
We know that disorder induces Anderson localization
\cite{Anderson58}. It is thus fundamental to ask whether the
localization near half-filling of cuprates is due to extrinsic
disorder effect\cite{Kastner98} or intrinsic quantum effect of the
single-band $t$-$J$ model\cite{Weng01}.
In other words, is the translation symmetry at the ground state of the $t$-$J$ model
\emph{spontaneously} broken?

Toward this direction, new progress has been made recently. Z. Zhu \emph{et al.} perform a large-scale density matrix
renormalization group (DMRG) numerical simulation by keeping enough
number of  states in each DMRG block with high accuracy.\cite{Zhu} 
The numerical result demonstrates that the density of single hole is localized in ladder systems,  
which was  explained by the so-called {\it phase string} theoretical argument (More details and analysis can be found in Ref. \onlinecite{Weng01}). Experimentally, C. Ye \emph{et al.}
\cite{STM} are able to significantly enlarge bias range in scanning
tunneling microscopy (STM) to study the atomic scale electronic
structure of the Ca$_2$CuO$_2$Cl$_2$ parent compound with
electron-donated defects. A well-defined in-gap state appears near
the edge of upper Hubbard band (UHB) and is strongly
localized in which the typical localization length is order of lattice constant.\cite{STM} These new findings combined with previous
experimental hints call for a coherent theoretical description on
charge dynamics of the $t$-$J$ model near half-filling which is the main
purpose of this work.

\begin{figure}[tbp]
\begin{center}
\includegraphics[width=8.8cm]{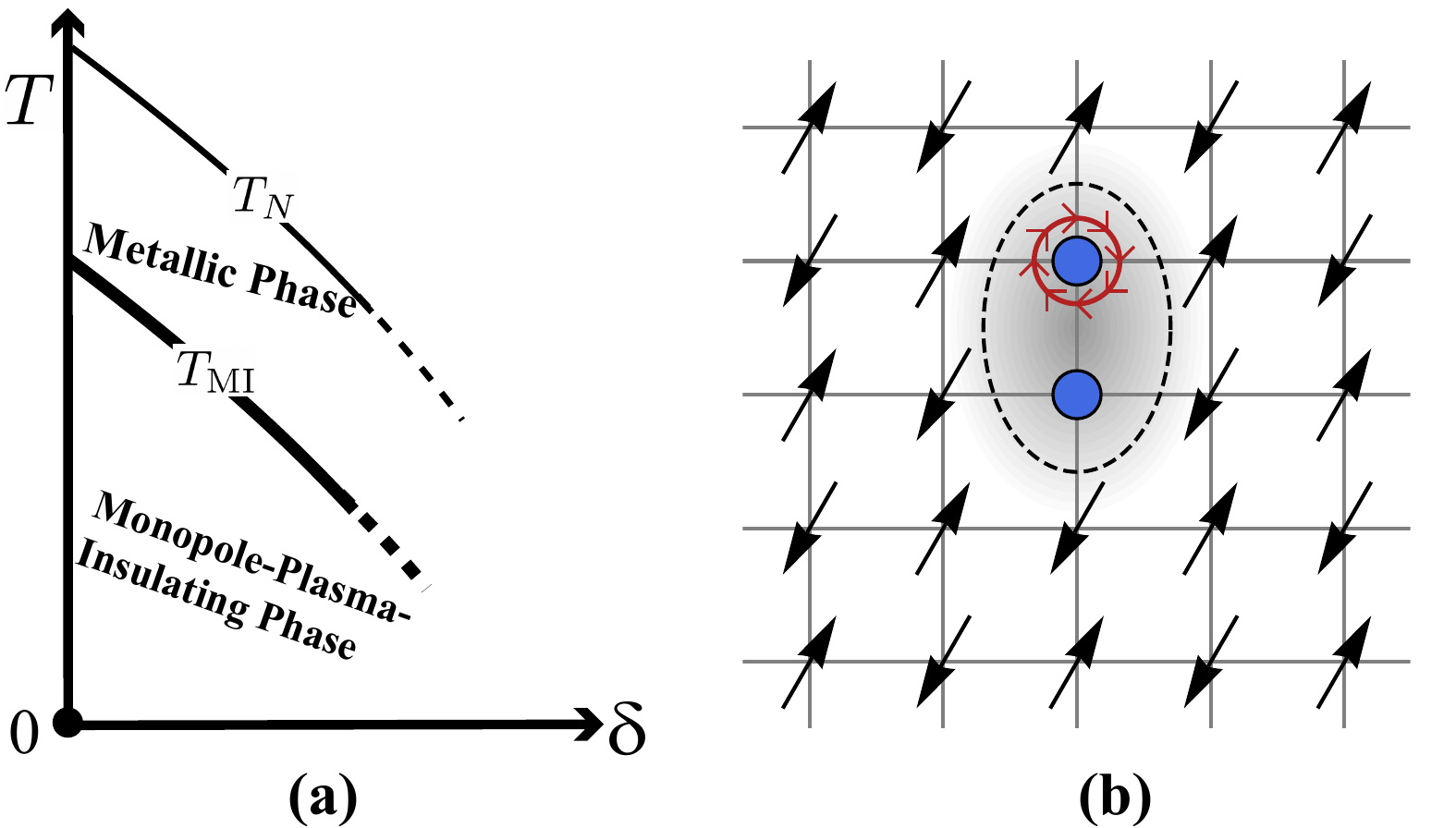}
\end{center}
\caption{(Color online) (a) Intrinsic phase diagram near
half-filling in absence of extrinsic disorder. $T$ and $\delta$
stand for temperature and hole concentration, respectively. The
phase region of $T<T_N$ supports antiferromagnetic long range order
(AFLRO). In this spin-ordered region, there is an additional
temperature scale $T_{\rm MI}$ about which the system undergoes
metal-insulator transition. The high (low)-$T$ region is metallic
phase (``Monopole-Plasma-Insulating Phase'') as to be explained in
main texts.   The dashed line segments imply that the present
analysis is in fact self-consistent only near half-filling. (b) A
schematic description of holon localization in AF background. Black
arrows form AF spin background and blue balls are two holons. The
red directional circle stands for a ``-2$\pi$'' phase-vortex whose
core is occupied by a holon. The three objects inside the dashed
circle form a localized bound state. } \label{figure_hole}
\end{figure}

In this paper, we shall provide a quantum
field-theoretic approach to show that charge localization is
driven by the non-perturbative monopole effect of a compact U(1)
gauge dynamics that was initially introduced in Polyakov's seminal
papers\cite{Polyakov75}.
We shall further propose the finite
temperature phase diagram (Fig. \ref{figure_hole}-a) near
half-filling in which the metal-insulator transition is identified
to monopole-driven confinement-deconfinement transition
characterized by the order parameter
``Polyalov-line''\cite{Polyakov78,tHooft78,Susskind79}.
Essentially, electrons are fractionalized into bosonic holons (carrying charge) and bosonic
spinons (carrying spin), both of which are mutually entangled via a mutual
Chern-Simons action.\cite{Ye1,Ye2} Spinons are condensed in
Arovas-Auerbach formalism\cite{Arovas88PRB} and form superfluid at
extremely low doping at zero temperature. At finite temperatures we
assume spinons are still condensed (by adding a weak interlayer AF
coupling $J_{\perp}$ or in-plane anisotropy $\alpha_{xy}$ in spin
space\cite{Chakravarty88,Hasenfratz91,Beard98,Keimer92,Kastner98})
with a nonzero N\'{e}el magnetic transition temperature $T_N$ where
Bose condensation occurs. (see Fig. \ref{figure_hole}-a) In such a
saddle point ansatz, we study charge dynamics at ground state and
finite temperature behavior below $T_N$.


First, the effective theory of charge dynamics is a (2+1)D compact
U(1) gauge theory coupled to holons and $\pm2\pi$ phase-vortices
arising from spinon superfluid. Since the gauge dynamics is in
confined phase where monopole operators are relevant, holons and
$\pm2\pi$ phase-vortices are enforced to form gauge-neutral bound
states.
Because phase-vortices are static particles with infinite effective mass, the bound states eventually localize even
without extrinsic disorder taken into account. Hence, the
translation symmetry is indeed broken spontaneously.

Second, a confinement-deconfinement transition of charge degree of
freedom occurs at a finite temperature $T_{\rm MI}$. At  $T<T_{\rm
MI}$, holons are confined via linear potential and localized such
that charge transport is of insulating nature. This low-temperature
phase is called  ``\emph{Monopole-Plasma-Insulating Phase}'' as to
be explained below. At $T>T_{\rm MI}$, the linear confinement
disappears, but holons still perceive a logarithmic interaction from
static phase-vortices of spinon superfluid background. Therefore,
this high-temperature metallic phase supports a metallic behavior.
$T_{\rm MI}$ is identified to the metal-insulator  transition
temperature scale observed in in-plane resistivity measurement of
heavily underdoped cuprates.\cite{Imada98,Ando01,Kastner98}   The
phase diagram is shown in Fig. \ref{figure_hole}-a. Disorder is
absent in the present theoretical approach but physically disorder effect in a realistic
material is expected to further amplify localization and thereby enhance
$T_{\rm MI}$.

The paper is organized as follows. In Sec. \ref{reviewsection}, the compact mutual Chern-Simons gauge theory of the $t$-$J$ model is reviewed where frequently used notations are introduced. The charge localization shall be derived in Sec. \ref{monopolesection} and the discussion of finite temperature phase diagram is arranged in Sec. \ref{temperaturesection}. All results are summarized in Sec. \ref{summarysection}.

\section{Compact mutual Chern-Simons theory of doped Mott insulators}\label{reviewsection}
In this section, we first review our understanding of doped Mott insulators,
especially the phase string effect  of the $t$-$J$ model and its compact mutual-Chern-Simons gauge theory.  More details can be found in Refs. \onlinecite{Weng95,Weng97,Kou05,Ye1,Ye2,Ye3}. We stress that the present $t$-$J$ model only contains nearest-hopping term and indeed the electron- and hole-doped cases are symmetric although the realistic cuprates have asymmetric phase diagram. By considering more hopping terms and super exchange terms, it is possible that the asymmetric phenomena can be realized which is beyond our present theoretical analysis.

Let's start with the so-called ``sign structure'' of doped spin model. The significance of figuring out the sign structure of a given theory can be
illustrated by the Nagaoka state\cite{Nagaoka},
one of the few exact results about the $t$-$J$ model.
The Nagaoka problem, the $U=\infty$ Hubbard model with one hole, is equivalent to
the large $J$ limit of Kondo lattice model with one conduction electron\cite{Kondo}.
It can be shown that the validity of Nagaoka theorem relies heavily
on the triviality of the sign structure of one hole hopping term of
the $t$-$J$ model\cite{Kondo}.
The sign structure of a model is called trivial when the Hamiltonian matrix has
only non-positive off-diagonal elements in some well-chosen basis.
If a Hamiltonian has trivial sign structure and satisfies some other conditions,
we can use Perron-Frobenius theorem to show that
the ground state of the system is non-degenerate and has a positive wave function
in this well-chosen basis.
It is the trivial sign structures of the AF Heisenberg model and the one-hole
infinity-$U$ Hubbard model that give rise to the Marshall theorem and the Nagaoka theorem.

At finite doping and finite $J$, however, the sign structure of the $t$-$J$ model is
highly nontrivial\cite{Wu08}.
We can also choose the Marshall bases $\{|\phi\rangle\}$
to make the off-diagonal matrix elements of Heisenberg term non-positive.
Nevertheless, nontrivial signs appear in the hopping term,
resulting in a model where the Perron-Frobenius theorem is no longer applicable.
This can be seen easily in slave-fermion formalism, where the hopping and Heisenberg terms
can be written as\cite{Weng97}
\begin{eqnarray}\label{Ht}
    H_t &=& -t\sum_{\langle i,j\rangle,\sigma} \sigma f_i^\dagger f_j
        b_{j\sigma}^\dagger b_{i\sigma}+\text{h.c.,}\\\label{HJ}
    H_J &=& -\frac{J}{2}\sum_{\langle i,j\rangle,\sigma,\sigma'}
        b_{i\sigma}^\dagger b_{j-\sigma}^\dagger b_{j-\sigma'}b_{i\sigma'}.
\end{eqnarray}
Here, $b_{i\sigma}$ is the annihilation operator of bosonic spinon at site $i$
with spin $\sigma$ ($\sigma=\pm$ denote spin up/down),
and $f_{i}$ the annihilation operator of fermionic holon at site $i$.
The configurations with more than one electrons at a certain site are projected out,
leaving a Hilbert space with constraint
\begin{equation}\label{constraint}
  \sum_\sigma b_{i\sigma}^\dagger b_{i\sigma} + f_i^\dagger f_i =1.
\end{equation}
The electron operator is written in this formalism as
\begin{equation}\label{elec}
  c_{i\sigma}=(-\sigma)^i f_i^\dagger b_{i\sigma},
\end{equation}
where the Marshall sign $(-\sigma)^i$ depends on the sublattice index ($i$ even or odd)
and the spin $\sigma$ of the electron.
The sign structure of the Heisenberg term Eq. (\ref{HJ}) is trivial,
while the spin dependent sign structure of hopping term Eq. (\ref{Ht}) indicates that
a minus sign appears whenever a down spinon exchanges with a holon.
This is called the \emph{phase string effect}\cite{Weng97}.

To keep track of this this nonrepairable phase string effect,
we define a nonlocal unitary transformation
\begin{equation}\label{nonlocalU}
  e^{i\hat\Theta} \equiv \exp\left(-i\sum_{i,l}\theta_i(l)n_i^h n_{l\downarrow}^b\right),
\end{equation}
where $\theta_i(l)=\text{Im}\ln(z_i-z_l)$,
and $z_l=x_l+iy_l$ is the complex coordinate of site $l$.
Under this nonlocal unitary transformation
$\hat O\rightarrow e^{i\hat\Theta}\hat O e^{-i\hat\Theta}$, the 
$t$-$J$ model becomes
\begin{eqnarray}\label{Ht2}
    H_t &=& -t\sum_{\langle i,j\rangle,\sigma} e^{iA_{ij}^s} h_i^\dagger h_j
        e^{i\sigma A_{ji}^h}b_{j\sigma}^\dagger b_{i\sigma}+\text{h.c.,}\\\label{HJ2}
    H_J &=& -\frac{J}{2}\sum_{\langle i,j\rangle,\sigma,\sigma'}
        e^{i\sigma A_{ij}^h} b_{i\sigma}^\dagger b_{j-\sigma}^\dagger
        e^{i\sigma' A_{ji}^h} b_{j-\sigma'}b_{i\sigma'}.
\end{eqnarray}
We have also used the Jordan-Wigner transformation in 2D to
describe holons by bosonic operators $h_i$'s instead of fermionic operators $f_i$'s.
The sign structure of the original $t$-$J$ model
after the nonlocal unitary transformation is now represented
by two compact U(1) gauge fields $A^s$ and $A^h$ defined by
\begin{eqnarray}\label{As}
    A_{ij}^s &\equiv& \frac{1}{2}\sum_{l\neq i,j}
        \left[\theta_i(l)-\theta_j(l)\right]
        \left(n_{l\uparrow}^b-n_{l\downarrow}^b\right)\ \ (\text{mod}\ 2\pi), \\\label{Ah}
    A_{ij}^h &\equiv& \frac{1}{2}\sum_{l\neq i,j}
        \left[\theta_i(l)-\theta_j(l)\right]n_l^h\ \ (\text{mod}\ 2\pi).
\end{eqnarray}
Those equations indicate that the holons (spinons) are the $\pi$ ($\pm\pi$) vortices of
the gauge field $A^h$ ($A^s$).

We can now use mean field theory to deal with this model,
since the phase string effect is explicitly tracked by
the nonlocal unitary transformation Eq. (\ref{nonlocalU}).
This procedure bears a resemblance to the simple direct product variational wave function
treatment of the Haldane phase of 1D AF spin-1 chain
after a nonlocal unitary transformation, which makes it possible to use local ferromagnetic
order parameter to reveal the nonlocal hidden string order of the phsse\cite{Haldane}.
The validity of the above method can be also illustrated by the 1D $t$-$J$ model which
possesses non-Fermi-liquid behavior.
In 1D case, $\theta_{i}(l)=\pi\cdot\theta(l-i)$,
where $\theta(x)$ is the Heaviside step function.
According to the definitions Eq. (\ref{As}) and (\ref{Ah}),
the gauge fields $A_{ij}^s$ and $A_{ij}^h$ vanish.
Thus the two terms in the $t$-$J$ model Eq. (\ref{Ht2}) and Eq. (\ref{HJ2})
both have a trivial sign structure, with all the signs absorbed into the definition of
the fractionalization of physical electron operator.
The simple mean field treatment is then enough to get the correct Luttinger-liquid behavior
of correlation functions\cite{Weng97}.

At mean field level, the $t$-$J$ model is reduced to the effective phase string model,
with Hamiltonian $H_{\text{eff}}=H_h+H_s$, and
\begin{eqnarray}
    H_h &=& -t_h \sum_{\langle i,j\rangle} e^{iA_{ij}^s} h_i^\dagger h_j
        +\text{h.c.,} \\
    H_s &=& -J_s \sum_{\langle i,j\rangle,\sigma}
        e^{i\sigma A_{ij}^h} b_{i\sigma}^\dagger b_{j-\sigma}^\dagger+\text{h.c.}\,,
\end{eqnarray}
where, $t_h$ and $J_s$ will be defined in below. This model possesses a compact U(1)$\otimes$U(1) gauge symmetry:
\begin{eqnarray}
    h_{i} &\rightarrow& h_{i}e^{i\theta_i^s}, \\
    b_{i\sigma} &\rightarrow& b_{i\sigma}e^{i\sigma\theta_i^h}, \\
  A_{ij}^{s,h} &\rightarrow& A_{ij}^{s,h}+\theta_i^{s,h}-\theta_j^{s,h}.
\end{eqnarray}

It is instructive to formulate the above mean field theory in path integral formalism after regularizing appropriately on a cubic spacetime lattice. This is the so-called mutual Chern-Simons gauge theory\cite{Ye1,Ye2}. Each holon (spinon), as a $\pi$ ($\pm \pi$)-flux tube, constitutes  ``electromagnetic'' flux of $A_\mu^h$ ($A_\mu^s$) which is minimally coupled to spinon (holon). This structure leads to compact mutual Chern-Simons topological term in which the compactness of $A_\mu^{s,h}$ is exactly protected. In such a formalism, the partition function encompasses compact U(1)$\otimes$U(1) gauge redundancy:
\begin{equation}
\mathcal{Z}=\sum_{\{ \mathscr{N}^{s}\},\{\mathscr{N}^{h}\}}\int
D[A^{s},A^{h}]D[h^{\dagger },h,b^{\dagger },b]e^{-S} \, \label{partition}
\end{equation}
in which the action $S=\sum_{x}\mathcal{L}$ ($x$ denotes spacetime
coordinates) with
$\mathcal{L}=\mathcal{L}_{h}+\mathcal{L}_{s}+\mathcal{L}_{\mathrm{MCS}}$.
$\mathcal{L}_{h}$ ($\mathcal{L}_{s}$) is the
Lagrangian density for holon (spinon), which includes usual gauge covariant operators responsible for \emph{minimal coupling} between holon (spinon) and $A^s$ ($A^h$). For further derivation, we need to explicitly write down the concrete form of $\mathcal{L}_s$:
\begin{align}
\mathcal{L}_{s}=&b_{i\sigma }^{\dagger }\left( d_{0}-i\sigma
A_{0}^{h}+\lambda ^{s}\right) \,b_{i\sigma }+\frac{u_{2}}{2}\left(
 b_{i\sigma }^{\dagger }b_{i\sigma }\right) ^{2}\nonumber\\
&-J_{s}\left( e^{i\sigma A_{\alpha }^{h}}b_{i+{\hat{\alpha}}\sigma }^{\dagger
}b_{i-\sigma }^{\dagger }+\text{\textrm{h.c.}}\right) \,,
\end{align}
where, Einstein summation is employed for all indices. $\alpha,\beta,\cdots$ denote space directions $\hat{x}$ or $\hat{y}$. The indices $\mu,\nu,\lambda\cdots$ stand for spacetime directions.  $d_\nu$ is forward difference operator on 3-dimensional spacetime lattice. $u_2$ is onsite repulsion energy which softens hard-core condition of Schwinger bosons.  $\lambda^s$ is chemical potential. The effective AF superexchange energy $J_s$ is defined as: $J_s=\frac{1}{2}\Delta^s$ where bosonic resonating-valence bond (RVB) order parameter $\Delta ^{s}\equiv \sum_{\sigma
}\langle e^{i\sigma A_{\alpha }^{h}}b_{i+{\hat{\alpha}}\sigma }^{\dagger
}b_{i-\sigma }^{\dagger }\rangle \neq 0$ below the pseudogap temperature $T_0$\cite{Gu}. This meanfield RVB condensate gaps out the usual  U(1) gauge fluctuation (denoted as $\text{U(1)}_A$) which arises from  single-occupancy constraint, in contrast to the strong gauge fluctuation in $\text{U(1)}_A$ slave-boson theory\cite{Lee06}. In the latter, it is necessary to carefully investigate confinement-deconfinement of the $\text{U(1)}_A$ gauge fluctuation by integrating out high energy modes of matter fields\cite{Lee06,Ichinose}. The compactified mutual Chern-Simons term is expressed by:
\begin{align}
\mathcal{L}_{\rm MCS}=\frac{i}{\pi }\, \epsilon ^{\mu \nu \lambda
}\left( A_{{\mu }}^{s}-2\pi \mathscr{N}_{\mu }^{s}\right) d_{\nu }\left( A_{{
\lambda }}^{h}-2\pi \mathscr{N}_{\lambda }^{h}\right)\,,
\end{align}
where, $\epsilon^{\mu\nu\lambda}$ is antisymmetric tensor of rank-3. $\mathscr{N}^{s,h}_\mu$ are two integer-valued link variables to take account of the periodicity of the gauge field ${A}_\mu^{s,h}$.

\section{Monopoles, Confinement and Localization}\label{monopolesection}
\subsection{Emergent compact gauge theory}
In this section, we focus on the AF phase in the formalism of mutual Chern-Simons theory. A compact U(1) gauge field theory emerges in the limit of low hole concentration.

At half-filling, spinon condensation leads to AFLRO. Near half-filling, we still  assume spinon condensation, i.e. $\langle b_\sigma\rangle\neq 0$. Let's formally write spinon field $b_\sigma= (\sqrt{n_0}+h)e^{i\sigma\theta}$ where $n_0$ is $\delta$-dependent condensation fraction. By integrating out the massive amplitude fluctuation $h$ of spinon field\cite{Zee}, one can obtain the following effective Lagrangian:
\begin{align}
  \mathcal{L}=&\frac{g_0}{2}(A^h_0-d_0\theta)^2+\frac{g_\alpha}{2}(A^h_\alpha-d_\alpha\theta)^2\nonumber\\
  &+\frac{i}{\pi}\epsilon^{\mu\nu\lambda}
  (A^s_\mu-2\pi  \mathscr{N}^s_\mu)d_\nu(A^h_\lambda-2\pi\mathscr{N}^h_\lambda)+\mathcal{L}_h\,,\label{eq_l}
\end{align}
where, $g_0={J_s}/{4}$, $g_1=g_2=4n_0J_s$.  $\mathscr{N}^h_\alpha$ plays the key role of the static $\pm 2\pi$ phase-vortices arising from the spinon superfluid. $\theta$ is a scalar function of spacetime coordinate. By further employing the unitary gauge\cite{Peskin}, $\theta$ can be absorbed into $A^h_\mu$,  while, $\mathcal{L}_{\rm MCS}$ keeps invariant due to the antisymmetry property of $\epsilon^{\mu\nu\lambda}$. Since $d_\mu\theta\in\mathbb{R}$, one obtains a massive real vector field $A^h_\mu$ with $A^h_\mu\in\mathbb{R}$.

Therefore, the first two terms in Eq. (\ref{eq_l}) are replaced by `` $\frac{g_0}{2}(A^h_0)^2+\frac{g_\alpha}{2}(A^h_\alpha)^2$'' by keeping $A^h_\mu\in\mathbb{R}$ in mind.
Gaussian integration over $A^h_\mu$ leads to:
\begin{align}\label{eq_com}
  \mathcal{L}=&\frac{1}{4\tilde{e}^2}(F^s_{\mu\nu}-2\pi n_{\mu\nu})^2+\mathcal{L}_c\,,
\end{align}
where, $\tilde{e}=\sqrt{4n_0 J_s}$   is coupling constant of
``emergent (2+1)D compact gauge dynamics''. The ``speed of light''
is implicit here without loss of generality. The gauge field
strength tensor $F^s_{\mu\nu}$ is defined as $F^s_{\mu\nu}=\hat{d}_\mu A^s_\nu-\hat{d}_\nu
A^s_\mu$ ($\hat{d}_\mu$ is backward difference operator on
3-dimensional spacetime lattice). The new plaquette variable
$n_{\mu\nu}$ is defined as: $n_{\mu\nu}\equiv (\hat{d}_\mu \mathscr{N}^s_\nu-\hat{d}_\nu
\mathscr{N}^s_\mu)$.
$\mathcal{L}_c=-i2\epsilon^{\mu\nu\lambda}A^s_\mu d_\nu
\mathscr{N}^h_\lambda+\mathcal{L}_h$. The definition of $n_{\mu\nu}$ here is locally well-defined. As will be discussed below,
$n_{\mu\nu}$  induces monopole configurations of three-dimensional
Euclidean spacetime when $\mathscr{N}^s_\mu$ has singularity. It is particularly interesting  that in the
quantum paramagnetic phase of non-linear sigma model, compact gauge
degree of freedom is also found and plays an important role in
classifying quantum spin
liquids.\cite{Haldane88,Senthil04,Balents07,Balents10} In these
systems, the compact gauge degree of freedom is essentially due to
the Wess-Zumino-Witten topological term of quantum SU(2)
spins.\cite{Haldane88}

It is clear that two kinds of particles simultaneously carry the gauge-charge of $A^s$, namely, holon $h$ in $\mathcal{L}^h$ and static $\pm 2\pi$ phase-vortex arising from spinon superfluid. Let's define (2+1)D current of phase-vortices as: $
 \epsilon^{\mu\nu\lambda}d_\nu \mathscr{N}^h_\lambda=  \mathscr{J}^\mu$,
where, $\mathscr{J}_0\in\mathbb{Z}$, $\mathscr{J}_\alpha=0$ (static vortices). As such, $\mathcal{L}_c$ can be simplified to
\begin{align}
\mathcal{L}_{\rm c}=-i A^s_0( 2\mathscr{J}_0+  h^\dagger h)+\mathcal{L}_{h\,[A^s_0=0]}\,,
\end{align}
where, $\mathcal{L}_{h\,[A^s_0=0]}$ stands for $\mathcal{L}_h$ without $A^s_0$ term. Hence, we find that the gauge-charge of $\mathscr{J}_0$ is $\pm2$ while each holon carries $+1$ gauge-charge, such that, only possible negative gauge-charge comes from $\mathscr{J}_0$.


\subsection{Monopole plasma configuration}

Monopole effect generically gets suppressed by finite density of matter field. Here, the holon matter field which couples to $A^s$ is extremely dilute, such that the monopole effect is
expected to be relevant.\cite{remark}
Let's briefly follow Polyakov's approach\cite{Polyakov75} by using the present mathematical symbols and explicitly keep track of monopole effect in context of doped antiferromagnets. In analog to the point-like solution ``Dirac monopole'' in three-dimensional space, we can define the ``magnetic field'' vector as below (in dual lattice): $\mathscr{B}_\mu=\frac{1}{2}\epsilon^{\mu\nu\lambda}n_{\nu\lambda}\cdot 2\pi$.
The divergence of $\mathscr{B}_\mu$ is in general quantized at $2\pi$, i.e., ${d}_\mu\mathscr{B}_\mu=2\pi q $
where $q$ is integer  scalar field defined on spacetime sites. $q\neq0$ if $\mathscr{N}^s_\mu$ has singularity. In general, $n_{\mu\nu}$  can be globally defined and  factorized into longitudinal and transversal components separately\cite{Polyakov75} (in the continuum limit, it becomes the Hodge decomposition of a general differential form into exact, co-exact and harmonic forms on a Riemannian manifold):
\begin{equation}\label{decomposition}
  n_{\mu\nu}= [\hat{d}_\mu(m_\nu+\chi_\nu)-\hat{d}_\nu(m_\mu+\chi_\mu) ]-\epsilon^{\mu\nu\lambda}\hat{d}_\lambda \phi,
\end{equation}
where, $m_\nu$ is integer vector field, $\chi_\nu$ is a real vector field with absolute value smaller than 1.  $\phi$ is a real  scalar field. It is a linear equation on lattice and one can check that the degrees of freedom of those fields on both sides of the equation are indeed the same. Substituting factorization formula of $n_{\mu\nu}$ into (\ref{eq_com}) we find that the $A^s_\mu$ can be combined with $m_\mu+\chi_\mu$ rendering $A^s_\mu\in \mathbb{R}$. This rearrangement brings convenience for the Gaussian integration for $A^s_\mu$. The effective Lagrangian is thus transformed into:
\begin{align}
\mathcal{L}=\frac{1}{4\tilde{e}^2}(F^s_{\mu\nu})^2-\frac{2\pi^2}{\tilde{e}^2}\phi {d}_\mu \hat{d}_\mu \phi+\mathcal{L}_c\,,\nonumber
\end{align}
where, $A^s_\mu\in\mathbb{R}$. In deriving above expression, the crossing term ``$F^s_{\mu\nu}\epsilon^{\mu\nu\lambda}\hat{d}_\lambda \phi$'' is neglected, because
\begin{align}
F^s_{\mu\nu}\epsilon^{\mu\nu\lambda}\hat{d}_\lambda \phi=&2\epsilon^{\mu\nu\lambda}\hat{d}_\mu{A}^s_\nu\hat{d}_\lambda\phi=2\epsilon^{\mu\nu\lambda}\hat{d}_\mu({A}^s_\nu\hat{d}_\lambda\phi)\nonumber\\&-
2\epsilon^{\mu\nu\lambda}{A}^s_\nu\hat{d}_\mu\hat{d}_\lambda\phi\,,\nonumber
\end{align}
where the first term is trivial boundary term and the second term vanishes due to antisymmetry property of $\epsilon^{\mu\nu\lambda}$.
To proceed further, let's substitute factorization formula of $n_{\mu\nu}$ into $d_\mu\mathscr{B}_\mu$ resulting in:
$-\hat{\Delta} \phi =  q\,$ where $\hat{\Delta}\equiv d_\mu\hat{d}_\mu$ is lattice Laplacian. The formal solution of $\phi$ can be written as: $
\phi=-\hat{\Delta}^{-1}q\,$.

Finally, we obtain the final effective action:
\begin{align}
\mathcal{S}_{\rm eff}=\sum_x\frac{(F^s_{\mu\nu})^2}{4\tilde{e}^2}-\frac{2\pi^2}{\tilde{e}^2}\sum_{x,x'} q_x (\hat{\Delta}^{-1})_{x,x'}q_{x'}+\sum_x\mathcal{L}_c\,,\label{action}
\end{align}
where, the second term in the above action describes a 3D plasma of monopoles   with Coulomb interaction $(-\hat{\Delta}^{-1})_{x,x'}\sim {|x-x'|}^{-1}$. The configuration $\{q\}$ represents  distribution of point-like ``magnetic charge'' (i.e. monopole).
This Coulomb gas (monopole plasma) representation of compact U(1) gauge theory serves as the starting point of our following discussions.

\subsection{Confinement and localization at zero temperature}
The monopole plasma has far-reaching consequences: the long-range interaction in monopole plasma spoils out the correlation of original gapless $A^s$-photon at weak-coupling limit and generates a gap in the low-lying charge excitation spectrum. This gap generation can be viewed as an alternative physical picture of Mott physics which strongly freezes charge degree of freedom near half-filling.

The matter field $h$ can be neglected at half-filling. In the absence of $\mathcal{L}_c$, Eq. (\ref{eq_com}) is identified to (2+1)D \emph{pure} compact U(1) gauge theory\cite{Polyakov75,Smit}, with Eq. (\ref{action}) as its monopole plasma representation.
To show the confining nature of the (2+1)D pure compact U(1) gauge theory at zero temperature, we can introduce a scalar field $\chi$ to reexpress the instanton part of the partition function. Taking into account only $q_x=\pm1$ configurations, we obtain the following sine-Gordon action:
\begin{equation}\label{sineGordon}
  \mathcal{S}_{\rm eff}=\left(\frac{\tilde{e}}{2\pi}\right)^2\sum_x \left((\nabla\chi)^2-M^2\cos\chi\right),
\end{equation}
where $M^2=(2\pi/\tilde{e})^2\exp(-\text{const.}/\tilde{e}^2)$. $\chi$ plays the role of scalar potential of the Coulomb charges, and its gradient is the electric (magnetic) field. The appearance of small mass $M$ of $\chi$ in weak coupling $\tilde{e}$ limit leads to a short range correlation function of electromagnetic field of the original U(1) theory. It is the finite density monopoles with long range interaction that spoil the correlations.

In order to probe confinement of gauge-charge, one can define Wilson loop\cite{Wilson74} as
\begin{align}
W[\mathcal{C}]=\langle e^{i\sum_x A^s_\mu J_\mu }\rangle\,,\nonumber
\end{align}
where, $\mathcal{C}$ is a temporal rectangular $r\times t$ loop with $r$ ($t$) spatial (temporal) distance. $J_\mu$ is unit current and forms the directional loop $\mathcal{C}$. The underlying potential between test-particle and test-antiparticle is defined as $V(r)=-\lim_{t\rightarrow\infty}\frac{\ln W[\mathcal{C}]}{t}$.
Repeating the same transformation from Eq. (\ref{eq_com}) to Eq. (\ref{action}), we find the similar formula for $W[\mathcal{C}]$:
\begin{align}
  W[\mathcal{C}]\sim& Z^{-1}\sum_{q_x}\exp\left(-\frac{2\pi^2}{\tilde e}\sum_{x,x'} q_x (\hat{\Delta}^{-1})_{x,x'}q_{x'}\right)\nonumber\\
  &\times \exp\left(2\pi i\sum_{x,x'} Q_x(\hat{\Delta}^{-1})_{x,x'}q_{x'}\right)\,.\label{W}
\end{align}
It describe monopole plasma with a fixed external monopole configuration $Q_x$. The screening of the external monopole configuration requires a free energy proportional to the area of the rectangle $\mathcal{C}$. In other words, the Wilson loop exhibits area law at large $t$ limit: $W[\mathcal{C}]\sim e^{-\kappa t\, r}\,$
with positive coefficient $\kappa$ for any given coupling constant $\tilde{e}$.
Since strong coupling limit is always a confinement state, we can reasonably draw the conclusion that (2+1)D pure compact U(1) theory is always confined at zero temperature.
As a results, all virtual particles that carry gauge-charge in the \emph{vacuum} of (2+1)D pure compact U(1) gauge theory must be confined into gauge-charge neutral bound state.

Near half-filling, the doped dilute holes may be
directly viewed as test-particles in the gedanken-experiment which
physically interprets Wilson loops\cite{Wilson74}, resulting in
linear confinement between holes and appropriate amounts of
phase-vortices $\mathscr{J}_0$ of spinon superfluid. The infinite
effective mass of the latter leads to strong localization of holes
(carrying charge degree of freedom). As shown in
Fig.\ref{figure_hole}-b, two holes form a localized state whose wave
function may be expressed as $|{\text{two holes}}\rangle
=|\vtop{\vskip-9pt\hbox{\includegraphics[width=12pt]{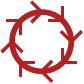}}},\vtop{\vskip-9pt\hbox{\includegraphics[width=12pt]{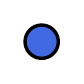}}},
\vtop{\vskip-9pt\hbox{\includegraphics[width=12pt]{holon.pdf}}}\rangle\,,$
where,
$\vtop{\vskip-9pt\hbox{\includegraphics[width=12pt]{vortex.pdf}}}$
and
$\vtop{\vskip-9pt\hbox{\includegraphics[width=12pt]{holon.pdf}}}$
denote $-2\pi$ phase-vortex and hole, respectively. It carries zero
gauge-charge (-2+1+1=0). As spin degree of freedom is energetically expelled away
from phase-vortex cores, a holon must occupy a core under the
single-occupancy constraint of the $t$-$J$ model, justifying Fig.
\ref{figure_hole}-b. Overall, we obtain the strong localization
without extrinsic disorder/impurities taken into account. Existence
of monopole effect in a spinful charge-neutral superfluid drives the
charge localization, and, the translation symmetry is broken
spontaneously.

Since the present quantum field-theoretic approach is based on the
phase string decomposition of electrons\cite{Sheng96,Weng97} as shown in Sec. \ref{reviewsection}, we emphasize
that the essential origin of localization mechanism can be traced
back to the singular phase string effect discovered by Weng \emph{et
al.}\cite{Sheng96,Weng97,Weng01}. The phase
string effect is mathematically captured by the exact ``sign
structure'' of the $t$-$J$ model by formulating partition function with
the worldline path-integral.\cite{Wu08} Pictorially, the worldlines
of one holon and one spin-$\downarrow$ spinon wrap each other can
contribute a minus sign under Gutzwiller projection (the total
particle number of spinons and holons at each site must always be
one). Such a particular sign structure encodes underlying non-Fermi
liquid behaviors and implies the notion of ``Sign Matter''
introduced by Zaanen and Overbosch.\cite{Zaanen11}

\section{Phase diagram of finite temperature}\label{temperaturesection}
\subsection{The effect of finite temperature}
At finite temperature, the compact U(1) theory has a deconfinement phase where the behavior of holons changes dramatically. We identify this confinement-deconfinement transition with the metal-insulator transition observed in experiments.

The same procedure dealing with zero temperature U(1) theory is valid at finite temperature. The two differences are: (i) Three dimensional infinite lattice is replaced by a lattice with imaginary time size $\beta=1/k_B T$. And only periodic configurations along this direction contribute to the partition function. (ii) When calculating the Green's function $G(x-x')=-(\hat\Delta^{-1})_{x,x'}$ in Eq.~(\ref{action}), the integral along the temporal direction is replaced by the Matsubara frequency summation. Using the fact that the Fourier transformation of the lattice Laplacian operator $\hat{\Delta}_{x,x'} = \sum_\mu ( \delta_{x,x'+\mu} + \delta_{x,x'-\mu} - 2 \delta_{x,x'})$ is $4\sum_\mu \sin^2(k_\mu/2)$, the Green's function in real space is given by
\begin{equation}\label{GreensFunction}
  G(\tau,\bold x)=\frac{1}{\beta}\int_{-\pi}^{\pi}\frac{\mathrm d^2\bold k}{(2\pi)^2}\sum_{n=-\infty}^{\infty}\frac{\exp(i\bold k\cdot\bold x+i\omega_n\tau)}{4\sum_i\sin^2\frac{k_i}{2}+4\sin^2\frac{\omega_n}{2}}.
\end{equation}
One can show that at zero temperature limit $\beta\rightarrow\infty$, the Green's function reproduces the three dimensional Coulomb potential at large distance
\begin{equation}
  G(\tau,\bold x)\sim \frac{1}{\sqrt{\bold x^2+\tau^2}},\quad \beta\rightarrow\infty.
\end{equation}
On the other hand, the high temperature limit $\beta\rightarrow 0$ give us an two dimensional Coulomb potential
\begin{equation}
  G(\bold x)\sim \frac{1}{\beta}\int_{-\pi}^{\pi}\frac{\mathrm{d}^2\bold k}{(2\pi)^2} \frac{\exp(i\bold k\cdot\bold x)}{4\sum_i\sin^2(k_i/2)},\quad \beta\rightarrow 0.
\end{equation}

We conclude that finite temperature effectively reduce one dimension of our original theory. The Coulomb gas has only one plasma phase in 3D; while in 2D, there is a Berezinskii-Kosterlitz-Thouless transition from disorder phase to critical phase. The sine-Gordon representation Eq. (\ref{sineGordon}) of the Coulomb gas is also valid. And the 2D sine-Gordon model with inverse temperature $\beta'=1/(\beta\tilde e^2)$  is a well studied model in conformal field theory. At low temperature (large $\beta$, small $\beta'$) of our original U(1) theory, the monopole effect term $\cos\chi$ is relevant, resulting in a short range correlation function of $\chi$, and the compact U(1) theory is in the confinement phase. At high temperature (small $\beta$, large $\beta'$), however, the monopole term is irrelevant, and the compact U(1) theory is in the deconfinement phase. The confinement-deconfinement transition do have measurable effect in our original $t$-$J$ model.

\subsection{Metal-insulator transition and order parameter}

Experimentally, magneto-resistance  measurement\cite{Lavrov99} of heavily underdoped YBa$_2$Cu$_3$O$_{6+x}$ compounds indicates that the development of AF order has little effect on the in-plane resistivity. It is also found that the magnitude of in-plane resistivity is so large that conventional band theory breaks down.\cite{Ando01,Emery95} In the present theory, gauge-charge is linearly confined at all $\tilde{e}$'s at zero temperature. According to Svetitsky-Yaffe universality arguments\cite{Svetitsky86}, it is suggested that a confinement-deconfinement transition exists at a finite temperature $T_{\rm MI}$. Therefore, it is naturally addressed that electric charge transport may be explicitly altered when the system undergoes the transition.
At $T<T_{\rm MI}$ holons are still linearly confined and strongly localized as same as ground state, implying insulating nature of charge transport. The only source for providing charge mobility is thermal fluctuation. At $T>T_{\rm MI}$, the linear confinement disappears such that the charge mobility is enhanced. Albeit disappearance of linear confinement, the logarithmic interaction now plays the leading role and renders a  metallic behavior. It is widely believed that the confinement-deconfinement transition temperature $T_{\rm MI}$ as a function of $\tilde{e}$ starts from origin $(0,0)$ in $T$-$\tilde{e}$ plane and roughly monotonically increases with the increase of $\tilde{e}$.\cite{Parga81,Coddington86,Svetitsky86,Chernodub01}
Doping holes in general depletes the spinon condensation fraction such that $n_0(\delta)$ decreases with the increase of doping $\delta$, so does the coupling constant $\tilde{e}$ which is defined as $\sqrt{4n_0J_s}$. Correspondingly, $T_{\rm MI}$ should monotonically decrease.
Consequently, we identify $T_{\rm MI}$ as the metal-insulator transition temperature scale observed in in-plane resistivity measurement.\cite{Ando01} The order parameter of this transition is so-called ``Polyakov-line''\cite{Polyakov78,tHooft78,Susskind79}:
\begin{align}
  \mathcal{O}\equiv \left\langle\exp\bigg\{ i\int^\beta_0 A^s_0 d\tau\bigg\} \right\rangle\,
\end{align}
with $\beta^{-1}={k_B T}$ and $k_B$ the Boltzmann constant.
$\mathcal{O}=(\neq)0$ if $T<(>)T_{\rm MI}$, which characterizes the
preservation (breaking) of central group. The Polyakov-line was ever
utilized in ``Short-Range Order phase'' of Hubbard model by
Wiegmann\cite{Wiegmann} but in quite a different context. Finally,
we obtain the phase diagram shown in Fig. \ref{figure_hole}-a.
Starting from the present understanding on localization,
quantitative study of charge transport is quite interesting and will
be extensively addressed in our future work.

\subsection{Dual-type ``Nernst'' effect: A new quantum phenomenon}
At the end of the discussion, we in particular emphasize the novel vortex-core structure near half-filling.  In Fig. \ref{figure_hole}-b, a quantized unit electric charge (i.e. holon, the blue ball) is surrounded by spinful supercurrent. For the lower holon, the supercurrent is counter-clockwise; for the upper holon, the existence of the extra $-2\pi$ phase-vortex (red directional circle) leads to net clockwise supercurrent. In analog to Nernst effect\cite{Wangyy06} in which vortex is formed by electric supercurrent, we predict that there is a dual-type effect if one can polarize spinful vorticity along $\hat{z}$-direction (cuprate sample is put in xy-plane). Then, by applying   temperature gradient along $\hat{x}$-direction, one can measure net spin accumulation at the two edges of $\hat{y}$-direction.

\section{Summary}\label{summarysection}
In conclusion, the present work provides a semi-quantitative quantum field-theoretic analysis on the long-standing problem: can electric charge be intrinsically localized in the $t$-$J$ model?. The monopole plasma configuration which comes from the antiferromagnetic background is proved to play fundamental role of driving charged holes into localized states. Although it is a technical challenge to determine the concrete range of parameter $t/J$ in which the result makes sense from the present semi-quantitative analysis, our work has proved  that the pure $t$-$J$ model itself has the \emph{intrinsic ability} to stabilize such translation symmetry breaking phase without the help of external disorder/impurity, which fundamentally differentiates the present localized ground state from Anderson localization. This result is a reasonable answer to the recent STM experimental finding and consistent to the phase string argument for DMRG numerical simulation of ladder systems.
Apart from this result, we also figure out the finite temperature phase diagram and especially theoretically explain the mechanism of metal-insulator transition found in electric in-plane resistivity measurement, which sheds light on a new way to reorganize transport experimental findings in curates in a single framework. A much more quantitative study along the present perspective is important and will be leaved to future work.

\section*{Acknowledgements}
We are especially grateful to Zheng-Yu Weng and Yayu Wang for enlightening discussions during preparation of the work. P.Y. would like to thank Meng Cheng, Su-Peng Kou, Joseph Maciejko and Yi-Zhuang You for their reading and helpful improvement on our draft. This work is supported by NSFC Grant No.
10834003, 11174174, and by MOST National Program for Basic Research
Grants No. 2009CB929402 and No. 2010CB923003. Research at Perimeter Institute is supported by the?Government
of Canada through Industry Canada and by the Province of Ontario through the Ministry of
Research and Innovation.

$^\dag$ Email: pye@pitp.ca


\begin{thebibliography}{99}
\bibitem{remark} Correspondingly, the monopole effect in superconducting phase is not relevant becuase of the large finite density of spinons which couple to Maxwell gauge field $A^h$ there. See Refs. \cite{Ye2,Ye3}.
\end{thebibliography}
\end{document}